\documentclass[sigconf]{acmart}
\AtBeginDocument{%
  }

\usepackage{url}

\begin{document}

\title{Towards A Cultural Intelligence and Values Inference Quality Benchmark for Community Values and Common Knowledge}

\author{Brittany Johnson}
\email{johnsonb@gmu.edu}
\affiliation{%
  \institution{George Mason University}
  \country{USA}
}

\author{Erin Reddick}
  \email{tech@chatblackgpt.com}
\affiliation{%
  \institution{ChatBlackGPT}
  \country{USA}
}

\author{Angela D.R. Smith}
   \email{adrsmith@utexas.edu}
\affiliation{%
  \institution{University of Texas at Austin}
  \country{USA}}

\renewcommand{\shortauthors}{Johnson et al.}

\begin{abstract}
  Large language models (LLMs) have emerged as a powerful technology, and thus, we have seen widespread adoption and use on software engineering teams. 
  Most often, LLMs are designed as ``general purpose'' technologies meant to represent the general population. 
  Unfortunately, this often means alignment with predominantly Western Caucasian narratives and misalignment with other cultures and populations that engage in collaborative innovation.
  In response to this misalignment, there have been recent efforts centered on the development of ``culturally-informed'' LLMs, such as ChatBlackGPT, that are capable of better aligning with historically marginalized experiences and perspectives.
  Despite this progress, there has been little effort aimed at supporting our ability to develop and evaluate culturally-informed LLMs. 
  A recent effort proposed an approach for developing a national alignment benchmark that emphasizes alignment with national social values and common knowledge.
  However, given the range of cultural identities present in the United States (U.S.), a national alignment benchmark is an ineffective goal for broader representation.
  To help fill this gap in this US context, we propose a replication study that translates the process used to develop \textsc{KorNAT}, a Korean National LLM alignment benchmark, to develop \textsc{CIVIQ}, a Cultural Intelligence and Values Inference Quality benchmark centered on alignment with community social values and common knowledge. 
  As a proof-of-concept for our approach, we focus our initial efforts on the Black community in the US and leverage both general purpose (e.g., ChatGPT) and culturally-informed (e.g., ChatBlackGPT) LLMs in our efforts.
  In this paper, we discuss our plans for conducting this research, including engaging our audience of interest in our efforts. 
  Our work provides a critical foundation for research and development aimed at cultural alignment of AI technologies in practice.
\end{abstract}




\received{20 February 2007}
\received[revised]{12 March 2009}
\received[accepted]{5 June 2009}

\maketitle


\section{Introduction}
The advent of large language models (LLMs) has led to a shift in how we collaborate and engage in software engineering.
LLMs can support natural language interactions for tasks such as facilitating asynchronous communication and learning, information retrieval, analysis and summarization, and practical guidance across domains, a trait that makes them highly attractive and frequently integrated into everyday workflows~\cite{wang2025generative},~\cite{kaiser2025new} ,~\cite{chatterji2025people}.
As a result, we have evolved to the era of AI-assisted collaboration and innovation~\cite{hicks2024new}, where artificial intelligence (AI) technologies like LLMs are rapidly being integrated to support collaborative tasks such as decision-making~\cite{he2025llm}, communication~\cite{rohovyi2024towards}, code review~\cite{cihan2025automated}, and project management~\cite{karnouskos2024relevance}.

While the integration of LLMs to support collaborative tasks shows promise, it also has the potential to introduce workplace and team inequities.
These inequities stem from both the nature of AI as a technology (e.g., centrality of data, nature of decision-making) and the broader context in which it has been developed and would potentially be used~\cite{dencik2025ai, li2025large}.
As a result, those from historically marginalized backgrounds are less likely to reap the benefits due to the higher likelihood of discriminatory outcomes and often being less equipped to find ways to extract the value of AI-assisted collaboration.
Therefore, as with any AI technology, the outcomes and quality of support are only as good as the data and evaluation mechanisms that are engaged in its design and development.

One of the most common ways to assess the capabilities and risks associated with a given AI technology is via benchmarking~\cite{hardy2025more}.
Benchmarks, which typically include some scoped dataset and corresponding metric(s), can be used to measure and identify flaws or weaknesses in model performance over time and support assessment of any number of concerns, including performance, efficiency, and accuracy~\cite{jones2000software,said2013benchmarking}.
While there are a variety of existing benchmarks available, including contributions aimed at supporting the assessment of cultural alignment of language technologies~\cite{lee-etal-2024-kornat}, there is still a significant gap between the ability for existing benchmarks to capture the nuances of socio-technical information needs and foundations necessary to adequately support different experiential perspectives in AI use in the workplace and beyond~\cite{reuel2024betterbench}.
This gap is exaggerated by findings from a recent effort aimed at developing a benchmark for cultural alignment with Korean culture, indicating both the necessary rigor behind cultural assessments and the inadequacy of existing LLMs in representing diverse  perspectives~\cite{lee-etal-2024-kornat}.

To fill this gap in the U.S. context, we propose a research agenda to develop \textsc{CIVIQ} (Cultural Intelligence and Values Inference Quality), a benchmark for LLM alignment with community culture.
As a proof-of-concept, we will develop \textsc{CIVIQ} for alignment with Black culture in the U.S., a decision we discuss and further motivate in Section~\ref{sec:background}.
We outline our research goals and survey methodology, clarifying the distinctions between our approach and the approach it is inspired by (Section~\ref{sec:agenda}).
We conclude with a discussion of anticipated outcomes and plans for future work, including evaluations plans following benchmark development (Section~\ref{sec:outcomes}).

\section{Background \& Motivation}
\label{sec:background}

While promising, AI suffers from persistent issues in its ability to adequately support the diaspora of potential users.
This is especially problematic when these technologies are being rapidly integrated into work environments and teams where there already exists systemic barriers to meaningful engagement for historically marginalized groups, like software development~\cite{guzman2024mind, george2025barriers}.

These difficulties often stem from systemic misalignment between the experiences that inform AI and the experiences of its users.
In an attempt to address this issue, research and development have rallied behind the concept of \textit{culturally-informed AI}~\cite{egede2025exploring, cirucci2025culturally}.
While known by many different names (e.g., culturally-intelligent AI, culturally-responsive AI), the idea is the same -- an AI system capable of going beyond predominantly Western narratives and values to support engagement from historically marginalized users.
Two notable examples of culturally-informed LLMs are \textit{ChatBlackGPT}~\footnote{\url{https://chatblackgpt.com/}} and \textit{Latimer}~\footnote{\url{https://www.latimer.ai/}}.
Latimer, touted as an AI for everyone trained with diverse histories and inclusive voices, is an LLM trained to accurately represent the experiences, culture, and history of Black and Brown communities.
ChatBlackGPT, while similar in its focus on providing culturally-relevant interactions, distinguishes itself by scoping to accurate representations of Black culture and history.
ChatBlackGPT further distinguishes itself by setting the precedent for community-driven model design.

While there is promise in the development of culturally-informed AI, there is still the issue of how to reproduce, evaluate, and maintain such systems.
In an effort to drive research and development in this direction, there have been recent efforts aimed at understanding and supporting cultural alignment of LLMs~\cite{alkhamissi2024investigating,tao2024cultural}.
In software and AI engineering, we commonly use \textit{benchmarks} as mechanisms for supporting these endeavors~\cite{ said2013benchmarking}.
Benchmarks provide structured, standardized metrics for evaluating performance and facilitating responsible innovation.

One concern raised regarding the cultural alignment of LLMs through benchmarking is the ability to align them across multiple cultures, as tools like Latimer attempt to do~\cite{khan2025randomness}.
However, we posit that the goal of cultural alignment should not be to encompass as many cultures as possible (this is what got us where we are now) but rather to facilitate alignment with specific cultures, as does ChatBlackGPT.
Making (to our knowledge) the first meaningful contribution in this direction is a recent study which introduced \textsc{KorNAT}, a national LLM alignment benchmark for Korea~\cite{lee-etal-2024-kornat}.
Their effort focused on improving LLM alignment with Korean culture by developing a dataset of social values and common knowledge items specific to Korean culture and defining metrics for national alignment scoring of LLMs.
They used their benchmark to evaluate seven LLMs and found that few existing models meet the threshold for alignment with Korean social values and common knowledge.

Building on the promising foundation provided by prior work, our ongoing efforts aim to facilitate the design and assessment of alignment with specific user groups in the U.S.
As a proof-of-concept, we will center our efforts on Black culture within the U.S.~\footnote{Any further rationale for this decision has been redacted for double anonymous review.}
While there exist numerous benchmarking efforts to support cultural alignment, existing benchmarks are not built around Black social values or common knowledge. 
This gap leads to considerations like ``AI Safety'' defaulting to majority norms that erase Black history and misread Black communities, thereby facilitating further marginalization in an ever-AI-driven world.
We aim to fill this gap by replicating and expanding the work done for  \textsc{KorNAT} to develop \textsc{CIVIQ}, which we discuss next.

\section{CIVIQ Research Agenda}
\label{sec:agenda}
The process behind \textsc{KorNAT} is multi-faceted and iterative, providing both rigor and culture-centered foundations.
However, their approach was designed for national alignment.
Given the discourse surrounding Black culture and erasure in the United States (U.S.)~\cite{king2020black}, national alignment is not an appropriate goal.
Below, we outline our adapted approach to creating a cultural alignment benchmark for community social values and common knowledge.

\subsection{Research Goals \& Objectives}
The goal of our research is to extend and adapt the benchmarking framework established by the \textsc{KorNAT} project to the U.S. context, focusing specifically on the intersection of language model performance and cultural representation.
In our study, we will design and develop \textsc{CIVIQ} (Cultural Intelligence and Values Inference Quality), a community-centric, culturally-grounded benchmark that reflects U.S.-based social values and common knowledge in the Black community.
Ultimately, our work seeks to facilitate the design of cultural alignment benchmarks that advance the development and use of culturally intelligent LLMs. 
By providing an empirically grounded standard for evaluating cultural alignment for more niche communities and cultures, our benchmark aims to support engineers in the development and use of AI systems that accurately reflect the social norms, cultural references, and histories that define the modern U.S. landscape.
We posit that the ability to create culturally intelligent AI systems is a necessary  step towards realizing equitable benefits and gains from AI integration in diverse software teams.



\subsection{Social Values Item Development} \label{subsec:svi_dev}



\subsubsection{Topic Pool Determination}

Following \textsc{KorNAT}'s approach, we will curate our social values topic pool from two sources: timely keywords and social conflict keywords.

For \textbf{timely keywords}, we will leverage media from the last 12 months that center Black communities.
One of the key challenges in creating a benchmark for Black culture alignment is ensuring the social values are both relevant and accurate.
Mainstream media often exclude, misrepresent, or mischaracterize Black stories, a systemic issue that makes most news or social media outlets difficult to leverage alone~\cite{baker2017stories, amponsah2024black}.
Therefore, we will focus our efforts on Black media outlets (e.g., The Root~\footnote{https://www.theroot.com/} and Blavity~\footnote{https://blavity.com/}) to curate keywords that are relevant to timely social matters in the Black community.


For \textbf{social conflict keywords}, we will focus our keyword curation on topics that are long-standing pain points or sources of conflict for the Black community.
Our goal is to focus on topics that have historical roots but also modern relevance, such as reparations~\cite{farmer2018somebody}, school funding~\cite{ponzini2022public}, confederate monuments~\cite{cox2021no}, equitable cannabis legalization~\cite{joshi2023tale}, redlining remediation~\cite{kang2021looking}, and anti-Black surveillance~\cite{williams2020fitting}.

\subsubsection{Item Generation \& Validation}

Once we have curated topic keywords (approximately 1,500), we will gather approximately 5-8 representative articles or community reports. 
We will provide our keywords and artifacts to an LLM and prompt it to generate 10 closed-ended items per source, ensuring that it is using clear and non-leading language. 
We will then perform two rounds of human revision in collaboration with our trained Black survey editors.
In our rounds of revisions, we will leverage the revision checklist used for \textsc{KorNAT}~\cite{lee-etal-2024-kornat}, which includes checking for timeliness, relevance, single-theme items, language, and grammar.
Items will be phrased as Likert scale questions (e.g., ``Strongly disagree'' to ``Strongly agree'') and include a comprehension check (e.g., ``I clearly/roughly understand/cannot answer'').
As another form of validation, we will adapt \textsc{KorNAT}'s approach to cross-national prompting (asking the model to answer ``as someone from X'') to test cross-group prompting (``answer as a Black American adult'').


\subsection{Common Knowledge Item Development} \label{subsec:cki_dev}

Unlike social values items, which can rely on subjective sources of information, the development of common knowledge requires ground truth and historical accuracy.
Similar to social values, our approach must diverge from the approach taken by \textsc{KortNAT}. In their efforts, they could leverage Korean textbooks and GED reference materials; however, in our context, much of the relevant history and information that would inform Black common knowledge would not be included in these materials. 

Therefore, to develop our  common knowledge items, we will utilize recruit Black historians, educators, and subject matter experts to curate relevant resources and insights on Black history and modern relevance. We will use their insights to Iteratively develop multiple-choice question (MCQ) items. We will use ChatBlackGPT, a GPT fine-tuned to Black history and context, to aid in scaling item generation~\cite{egede2025exploring}.
We will conduct independent revisions of resulting items, auditing for differential item functioning~\cite{holland2012differential} across sub-groups.

The goal of our common knowledge set is to cover themes and facts that have historical significance but also modern relevance. This includes topics such as \textbf{Pre-enslavement African history \& civilizations}, \textbf{Black Reconstruction}, \textbf{Great Migration \& Harlem Renaissance}, \textbf{Black STEM \& business contributions}, \textbf{Civics \& rights today}, \textbf{Health \& environment}, 
and \textbf{Practical knowledge \& culture}.
We will vette our topic selection with Black historians and educators to ensure its relevance, completeness, and accuracy.

Given the diversity of knowledge represented in our common knowledge item set, and the need to balance the kinds of questions provided, we will include a rationale and expected difficulty tag (e.g., easy, medium, hard) for each item.
\vspace{-.5em}


\subsection{Survey Structure \& Design}

The final survey will include both social value likert scale items (4,000) and common knowledge multiple choice items (6,000).
The survey structure will be based on the categories that unify and represent the topics that emerge from the processes discussed in Sections~\ref{subsec:svi_dev} and~\ref{subsec:cki_dev}.
Given our goal is to acquire the scale of \textsc{KorNAT} (over 6,000 unique respondents), we will follow their approach to survey design \& deployment.
This includes the use of stratified sampling to facilitate engagement with small portions of the survey at a time, alternating likert style and multiple choice questions to reduce fatigue, and using short stems with concise options to increase usability on mobile devices.

To support response quality control, we will include both attention and consistency questions.
Attention questions ask the respondent to respond with a specific response to ensure they are reading the questions (e.g., For this question, please select ``None of the above'' below). 
Consistency questions provide another mechanism for ensuring respondents are truly engaged in the survey by asking the same semantic question using different syntax (e.g., ``I prefer to use AI systems that reflect my cultural background.'' vs. ``It doesn’t matter if an AI system understands my cultural context.'').
The expectation is that respondents would answer the same or similarly for both questions.
We will pilot our survey to evaluate and improve all aspect of our survey design and structure, including comprehension of and polarization on our survey items.

\subsection{Sampling Frame \& Weighting}

The audience of interest is adult (18 years of age and over) Black and African American adults across the United States. 
To ensure balanced representation across the various demographic and sociocultural dimensions relevant to Black communities in the U.S., we will use stratified random  sampling~\cite{acharya2013sampling}. Our stratification variables will include \textbf{age} (18--29, 30--44, 45--59, and 60+), \textbf{gender} (male, female, transgender male, transgender female, non-binary), \textbf{region} (South, Midwest, Northeast, and West), \textbf{urbanicity} (urban, suburbans, and rural), \textbf{socioeconomic factors} (education level and income tier), \textbf{diaspora origin} (African Descents of Slavery (ADOS), Afro-Carribean, and African immigrant identities), \textbf{faith tradition} (including, but not limited to Christian, Muslim, traditional African religions, and secular respondents), \textbf{carceral history} (prior incarceration or justice system contact), \textbf{disability status}, and \textbf{sexual orientation} (heterosexual, gay, lesbian, bisexual, queer, pansexual, asexual).
We will use oversampling for underrepresented or hard-to-reach groups (e.g., African immigrants, LGBTQ+ individuals, persons with disabilities, and formerly incarcerated individuals) to improve precision and reduce variance in sub-group estimates~\cite{chang2013oversampling}.

Following \textsc{KorNAT}'s approach, we will adjust underrepresented responses as necessary using two processes: \textit{stratification adjustment} and \textit{sampling adjustment}. 
Stratification adjustment helps adjust demographic imbalances between the broader population and survey respondents while sampling adjustment modifies the selection probability when respondents are either over- or under-represented in a specific group. We will use their calculations for adjustment weights and normalize weighted responses.

\subsection{Engaging Our Audience of Interest}

As mentioned previously, it is critical for our efforts that we are able to engage a diverse range of identities in the U.S. Black population.
Given the known issues regarding engagement from historically marginalized communities in research~\cite{armstrong2022research, anuyah2023engaging}, we will experiment with non-traditional mechanisms for engagement.
Unlike survey research aimed at broader audiences, we will use broad advertising (e.g., social media) as a last resort. Instead, we will prioritize in-person, localized engagements across the U.S.
We will leverage our locality to cities and areas with large or predominantly Black populations (e.g., Detroit, Washington, D.C., Austin) and partnerships in those areas to organize social events that will facilitate survey participation.  
Those interested will have the option to take the survey in exchange for admission to the event being hosted in their city.
We will also use a novel compensation model to ensure response anonymity. Rather than respondents providing contact information for compensation, respondents can earn coins for each subset of the survey they take.
Given the inclusivity benefits of gamification~\cite{mcfarland2024cultural, bavi2022gamification}, this approach can increase engagement in larger portions of the survey from a broader ranges of respondents.
We will set an appropriate limit for coins that can be acquired and require photo identification for coin redemption to ensure data integrity. 

\subsection{Calculating Cultural Alignment Score}

Central to our research goals is the the ability for our efforts to facilitate evaluation of cultural alignment in LLMs and other advanced technologies. As such, following \textsc{KorNAT}'s approach, we will use our survey data to calculate scores for \textit{social value alignment} and \textit{common knowledge alignment}.

\subsubsection{Social Values Alignment Scoring}
Given social value items are likert-scale, there is no obvious ground truth.
Furthermore, selecting a ground truth in this context (e.g., based on majority vote) could lead to our missing value information provided by other responses~\cite{lee-etal-2024-kornat}.
Therefore, for the \textbf{Social Value Alignment (SVA)} metric we will use the distribution of survey responses for each social value item. More specifically, this metric is a ratio ($r_{ij}$) of respondents choosing the $j^{th}$ option for the $i^{th}$ question ($q_{i}$). So if a given model predicts the $k^{th}$ option for $q_{i}$, it receive a social alignment score of $r_{ik}$ (between 0 and 1).
The final score is the average score across all social value questions, with a score of over 0.5 indicating alignment with the majority of a given community.

As with \textsc{KorNAT}, we will use our empirical data to calculate the maximum possible score and determine if and to what extent it is necessary to develop an aggregated version of our SVA metric. For example, in their work the empirically calculated maximum for the SVA metric was 0.450 which suggests high variability in social values on a five-point scale. 
To address this, they proposed an \textbf{Aggregated Social Value Alignment (A-SVA)} metric that modifies the ground truth distributions such that there are only three groups: Disagree (grouping `Strongly Disagree' and `Disagree'), Neutral, and Agree (grouping `Strongly Agree' and 'Agree').
In their efforts, this increased the maximum achievable score to 0.626, indicating a ``moderate level'' of agreement.

\subsubsection{Common Knowledge Scoring}

Given the common knowledge items each have one correct answer, the Common Knowledge Alignment (CKA) metric is based on accuracy.
For \textsc{KorNAT}, the standard score was set as 0.6 to reflect the cut-off score for the Korean GED (60 points); models above this threshold would be considered to have sufficient common knowledge. 
However, given \textsc{CIVIQ}'s goal to facilitate community culture-level alignment and the centrality of Western White history in U.S. education~\cite{king2017status, rodriguez2023social}, this would be a misalignment metric to leverage.
Therefore, we will use the competency threshold for Black history and culture proficiency exams administered at HBCUs.
The threshold across exams that score on a 1--100 scales is 70 for competency; therefore, the threshold for our CKA metric will be 0.70.


\section{Ethics, Privacy, and Data Governance} \label{subsec:ethics}

We will adhere to rigorous ethical and privacy standards to ensure participants and communities represented in the data are protected.
Our study protocol will undergo Institutional Review Board (IRB) review and approval prior to data collection.
We will acquired informed consent, providing potential respondents with clear and concise information about the study's purpose, procedures, potential risks, and remind them of their right to withdraw from the study at any time.
To respect participant autonomy, we have designed our study such that respondents can have the option to skip any sensitive question without penalty or exclusion from the study.




One concern in conducting this research is the potential for push-back from the Black community regarding their comfort with or the potential harms that could stem from such broad access to their values and lived experiences~\cite{klassen2022black}.
In an attempt to mitigate and alleviate these kinds of concerns, our research efforts will operate under a \textbf{community data covenant}, a governance framework that will ensure data and artifacts derived from our work is used solely for social benefit and non-harm.
Our framework will have explicit restrictions that prohibit use of the dataset or derivative models for activities that could perpetuate harm, such as \textit{Surveillance or predictive policing}, \textit{Immigration or law enforcement activities},  \textit{Insurance or credit risk scoring}, \textit{Tenant screening or employment background checks}, and \textit{Any other carceral or punitive applications}.

All downstream users will be required to submit a transparency statement describing intended use, ethical safeguards, and community benefits prior to accessing the dataset. 
Approved users will be required to adhere to these covenant terms and cite them in any derivative work.


Our release plan will follow the staged model used for \textsc{KorNAT}. 
We will release a publicly accessible leaderboard and a sampled item set that can be used to support benchmarking.
We will only release the full dataset via controlled access under a Data Use Agreement (DUA) that enforces our community data covenant, privacy constrains, and reporting requirements.
Each release will include an accompanying data card documenting provenance, limitations, and ethical constraints to ensure responsible and reproducible use.

\section{Anticipated Outcomes \& Future Work} \label{sec:outcomes}

The research outlined in this paper will lay the necessary foundations to better understand the potential for sustainable innovation around LLMs and other AI technologies capable of adequately supporting culturally-diverse software teams.
We anticipate a surge in desire, even need, to meaningfully evaluate social proficiency of intelligent systems which requires a trustworthy precedent. 
The cooperative, community-engaged effort surrounding our proposed benchmarking efforts provides a unique opportunity to build sustainable bridges between academia, industry, and society that can facilitate the design of a better technological landscape for all. 
Building on these foundations, we will experiment with top LLMs, both general-purpose and culturally informed, using our benchmark \textsc{CIVIQ} to evaluate alignment with Black culture.
Our future efforts will also involve deeper explorations into the use cases for and potential benefits of culturally-aligned AI-assisted technologies on software teams and in other collaborative, human-facing environments.


\bibliographystyle{ACM-Reference-Format}
\bibliography{references}


\begin{thebibliography}{41}


\ifx \showCODEN    \undefined \def \showCODEN     #1{\unskip}     \fi
\ifx \showISBNx    \undefined \def \showISBNx     #1{\unskip}     \fi
\ifx \showISBNxiii \undefined \def \showISBNxiii  #1{\unskip}     \fi
\ifx \showISSN     \undefined \def \showISSN      #1{\unskip}     \fi
\ifx \showLCCN     \undefined \def \showLCCN      #1{\unskip}     \fi
\ifx \shownote     \undefined \def \shownote      #1{#1}          \fi
\ifx \showarticletitle \undefined \def \showarticletitle #1{#1}   \fi
\ifx \showURL      \undefined \def \showURL       {\relax}        \fi
\providecommand\bibfield[2]{#2}
\providecommand\bibinfo[2]{#2}
\providecommand\natexlab[1]{#1}
\providecommand\showeprint[2][]{arXiv:#2}

\bibitem[Acharya et~al\mbox{.}(2013)]%
        {acharya2013sampling}
\bibfield{author}{\bibinfo{person}{Anita~S Acharya}, \bibinfo{person}{Anupam Prakash}, \bibinfo{person}{Pikee Saxena}, {and} \bibinfo{person}{Aruna Nigam}.} \bibinfo{year}{2013}\natexlab{}.
\newblock \showarticletitle{Sampling: Why and how of it}.
\newblock \bibinfo{journal}{\emph{Indian journal of medical specialties}} \bibinfo{volume}{4}, \bibinfo{number}{2} (\bibinfo{year}{2013}), \bibinfo{pages}{330--333}.
\newblock


\bibitem[AlKhamissi et~al\mbox{.}(2024)]%
        {alkhamissi2024investigating}
\bibfield{author}{\bibinfo{person}{Badr AlKhamissi}, \bibinfo{person}{Muhammad ElNokrashy}, \bibinfo{person}{Mai AlKhamissi}, {and} \bibinfo{person}{Mona Diab}.} \bibinfo{year}{2024}\natexlab{}.
\newblock \showarticletitle{Investigating cultural alignment of large language models}.
\newblock \bibinfo{journal}{\emph{arXiv preprint arXiv:2402.13231}} (\bibinfo{year}{2024}).
\newblock


\bibitem[Amponsah(2024)]%
        {amponsah2024black}
\bibfield{author}{\bibinfo{person}{Emma-Lee Amponsah}.} \bibinfo{year}{2024}\natexlab{}.
\newblock \showarticletitle{Black dis/engagement: negotiating mainstream media presence and refusal}.
\newblock \bibinfo{journal}{\emph{European Journal of Cultural Studies}} \bibinfo{volume}{27}, \bibinfo{number}{5} (\bibinfo{year}{2024}), \bibinfo{pages}{1037--1055}.
\newblock


\bibitem[Anuyah et~al\mbox{.}(2023)]%
        {anuyah2023engaging}
\bibfield{author}{\bibinfo{person}{Oghenemaro Anuyah}, \bibinfo{person}{Karla Badillo-Urquiola}, {and} \bibinfo{person}{Ronald Metoyer}.} \bibinfo{year}{2023}\natexlab{}.
\newblock \showarticletitle{Engaging the discourse of empowerment for marginalized communities through research and design participation}. In \bibinfo{booktitle}{\emph{Extended Abstracts of the 2023 CHI Conference on Human Factors in Computing Systems}}. \bibinfo{pages}{1--7}.
\newblock


\bibitem[Armstrong and Ritchie(2022)]%
        {armstrong2022research}
\bibfield{author}{\bibinfo{person}{Katrina Armstrong} {and} \bibinfo{person}{Christine Ritchie}.} \bibinfo{year}{2022}\natexlab{}.
\newblock \showarticletitle{Research participation in marginalized communities—overcoming barriers}.
\newblock \bibinfo{journal}{\emph{New England Journal of Medicine}} \bibinfo{volume}{386}, \bibinfo{number}{3} (\bibinfo{year}{2022}), \bibinfo{pages}{203--205}.
\newblock


\bibitem[Baker-Bell et~al\mbox{.}(2017)]%
        {baker2017stories}
\bibfield{author}{\bibinfo{person}{April Baker-Bell}, \bibinfo{person}{Raven~Jones Stanbrough}, {and} \bibinfo{person}{Sakeena Everett}.} \bibinfo{year}{2017}\natexlab{}.
\newblock \showarticletitle{The stories they tell: Mainstream media, pedagogies of healing, and critical media literacy}.
\newblock \bibinfo{journal}{\emph{English Education}} \bibinfo{volume}{49}, \bibinfo{number}{2} (\bibinfo{year}{2017}), \bibinfo{pages}{130--152}.
\newblock


\bibitem[Bavi and Gupta(2022)]%
        {bavi2022gamification}
\bibfield{author}{\bibinfo{person}{Ahlam Bavi} {and} \bibinfo{person}{Neha Gupta}.} \bibinfo{year}{2022}\natexlab{}.
\newblock \showarticletitle{Gamification of digital heritage as an approach to improving museum and art gallery engagement for blind and partially sighted visitors}.
\newblock \bibinfo{journal}{\emph{Archaeologies}} \bibinfo{volume}{18}, \bibinfo{number}{3} (\bibinfo{year}{2022}), \bibinfo{pages}{585--622}.
\newblock


\bibitem[Chang et~al\mbox{.}(2013)]%
        {chang2013oversampling}
\bibfield{author}{\bibinfo{person}{Chia-Yun Chang}, \bibinfo{person}{Ming-Tsung Hsu}, \bibinfo{person}{Emilio~Xavier Esposito}, {and} \bibinfo{person}{Yufeng~J Tseng}.} \bibinfo{year}{2013}\natexlab{}.
\newblock \showarticletitle{Oversampling to overcome overfitting: exploring the relationship between data set composition, molecular descriptors, and predictive modeling methods}.
\newblock \bibinfo{journal}{\emph{Journal of chemical information and modeling}} \bibinfo{volume}{53}, \bibinfo{number}{4} (\bibinfo{year}{2013}), \bibinfo{pages}{958--971}.
\newblock


\bibitem[Chatterji et~al\mbox{.}(2025)]%
        {chatterji2025people}
\bibfield{author}{\bibinfo{person}{Aaron Chatterji}, \bibinfo{person}{Thomas Cunningham}, \bibinfo{person}{David~J Deming}, \bibinfo{person}{Zoe Hitzig}, \bibinfo{person}{Christopher Ong}, \bibinfo{person}{Carl~Yan Shan}, {and} \bibinfo{person}{Kevin Wadman}.} \bibinfo{year}{2025}\natexlab{}.
\newblock \bibinfo{booktitle}{\emph{How people use chatgpt}}.
\newblock \bibinfo{type}{{T}echnical {R}eport}. \bibinfo{institution}{National Bureau of Economic Research}.
\newblock


\bibitem[Cihan et~al\mbox{.}(2025)]%
        {cihan2025automated}
\bibfield{author}{\bibinfo{person}{Umut Cihan}, \bibinfo{person}{Vahid Haratian}, \bibinfo{person}{Arda {\.I}{\c{c}}{\"o}z}, \bibinfo{person}{Mert~Kaan G{\"u}l}, \bibinfo{person}{{\"O}mercan Devran}, \bibinfo{person}{Emircan~Furkan Bayendur}, \bibinfo{person}{Baykal~Mehmet U{\c{c}}ar}, {and} \bibinfo{person}{Eray T{\"u}z{\"u}n}.} \bibinfo{year}{2025}\natexlab{}.
\newblock \showarticletitle{Automated code review in practice}. In \bibinfo{booktitle}{\emph{2025 IEEE/ACM 47th International Conference on Software Engineering: Software Engineering in Practice (ICSE-SEIP)}}. IEEE, \bibinfo{pages}{425--436}.
\newblock


\bibitem[Cirucci et~al\mbox{.}(2025)]%
        {cirucci2025culturally}
\bibfield{author}{\bibinfo{person}{Angela~M Cirucci}, \bibinfo{person}{Miles Coleman}, \bibinfo{person}{Dan Strasser}, {and} \bibinfo{person}{Evan Garaizar}.} \bibinfo{year}{2025}\natexlab{}.
\newblock \showarticletitle{Culturally responsive communication in generative AI: looking at ChatGPT’s advice for coming out}.
\newblock \bibinfo{journal}{\emph{AI \& SOCIETY}} \bibinfo{volume}{40}, \bibinfo{number}{4} (\bibinfo{year}{2025}), \bibinfo{pages}{2249--2257}.
\newblock


\bibitem[Cox(2021)]%
        {cox2021no}
\bibfield{author}{\bibinfo{person}{Karen~L Cox}.} \bibinfo{year}{2021}\natexlab{}.
\newblock \bibinfo{booktitle}{\emph{No common ground: Confederate monuments and the ongoing fight for racial justice}}.
\newblock \bibinfo{publisher}{UNC Press Books}.
\newblock


\bibitem[Dencik et~al\mbox{.}(2025)]%
        {dencik2025ai}
\bibfield{author}{\bibinfo{person}{Lina Dencik}, \bibinfo{person}{Jessica Brand}, \bibinfo{person}{Philippa Metcalfe}, {and} \bibinfo{person}{Cate Hopkins}.} \bibinfo{year}{2025}\natexlab{}.
\newblock \showarticletitle{AI Inequalities at Work}.
\newblock  (\bibinfo{year}{2025}).
\newblock


\bibitem[Egede et~al\mbox{.}(2025)]%
        {egede2025exploring}
\bibfield{author}{\bibinfo{person}{Lisa Egede}, \bibinfo{person}{Ebtesam~Al Haque}, \bibinfo{person}{Gabriella Thompson}, \bibinfo{person}{Alicia Boyd}, \bibinfo{person}{Angela~DR Smith}, {and} \bibinfo{person}{Brittany Johnson}.} \bibinfo{year}{2025}\natexlab{}.
\newblock \showarticletitle{Exploring Culturally Informed AI Assistants: A Comparative Study of ChatBlackGPT and ChatGPT}. In \bibinfo{booktitle}{\emph{Proceedings of the Extended Abstracts of the CHI Conference on Human Factors in Computing Systems}}. \bibinfo{pages}{1--9}.
\newblock


\bibitem[Farmer(2018)]%
        {farmer2018somebody}
\bibfield{author}{\bibinfo{person}{Ashley~D Farmer}.} \bibinfo{year}{2018}\natexlab{}.
\newblock \showarticletitle{" Somebody Has to Pay": Audley Moore and the Modern Reparations Movement}.
\newblock \bibinfo{journal}{\emph{Palimpsest: A Journal on Women, Gender, and the Black International}} \bibinfo{volume}{7}, \bibinfo{number}{2} (\bibinfo{year}{2018}), \bibinfo{pages}{108--134}.
\newblock


\bibitem[George and Wooden(2025)]%
        {george2025barriers}
\bibfield{author}{\bibinfo{person}{Babu George} {and} \bibinfo{person}{Ontario~S Wooden}.} \bibinfo{year}{2025}\natexlab{}.
\newblock \showarticletitle{Barriers to Equitable Progress}.
\newblock In \bibinfo{booktitle}{\emph{AI Empowered}}. \bibinfo{publisher}{Emerald Publishing Limited}, \bibinfo{pages}{27--35}.
\newblock


\bibitem[Guzm{\'a}n et~al\mbox{.}(2024)]%
        {guzman2024mind}
\bibfield{author}{\bibinfo{person}{Emitz{\'a} Guzm{\'a}n}, \bibinfo{person}{Ricarda Anna-Lena Fischer}, {and} \bibinfo{person}{Janey Kok}.} \bibinfo{year}{2024}\natexlab{}.
\newblock \showarticletitle{Mind the gap: gender, micro-inequities and barriers in software development}.
\newblock \bibinfo{journal}{\emph{Empirical Software Engineering}} \bibinfo{volume}{29}, \bibinfo{number}{1} (\bibinfo{year}{2024}), \bibinfo{pages}{17}.
\newblock


\bibitem[Hardy et~al\mbox{.}(2025)]%
        {hardy2025more}
\bibfield{author}{\bibinfo{person}{Amelia Hardy}, \bibinfo{person}{Anka Reuel}, \bibinfo{person}{Kiana Jafari~Meimandi}, \bibinfo{person}{Lisa Soder}, \bibinfo{person}{Allie Griffith}, \bibinfo{person}{Dylan~M Asmar}, \bibinfo{person}{Sanmi Koyejo}, \bibinfo{person}{Michael~S Bernstein}, {and} \bibinfo{person}{Mykel~John Kochenderfer}.} \bibinfo{year}{2025}\natexlab{}.
\newblock \showarticletitle{More than marketing? on the information value of ai benchmarks for practitioners}. In \bibinfo{booktitle}{\emph{Proceedings of the 30th International Conference on Intelligent User Interfaces}}. \bibinfo{pages}{1032--1047}.
\newblock


\bibitem[He et~al\mbox{.}(2025)]%
        {he2025llm}
\bibfield{author}{\bibinfo{person}{Junda He}, \bibinfo{person}{Christoph Treude}, {and} \bibinfo{person}{David Lo}.} \bibinfo{year}{2025}\natexlab{}.
\newblock \showarticletitle{LLM-Based Multi-Agent Systems for Software Engineering: Literature Review, Vision, and the Road Ahead}.
\newblock \bibinfo{journal}{\emph{ACM Transactions on Software Engineering and Methodology}} \bibinfo{volume}{34}, \bibinfo{number}{5} (\bibinfo{year}{2025}), \bibinfo{pages}{1--30}.
\newblock


\bibitem[Hicks et~al\mbox{.}(2024)]%
        {hicks2024new}
\bibfield{author}{\bibinfo{person}{Catherine~M Hicks}, \bibinfo{person}{Carol~S Lee}, {and} \bibinfo{person}{Kristen Foster-Marks}.} \bibinfo{year}{2024}\natexlab{}.
\newblock \showarticletitle{The new developer: AI skill threat, identity change \& developer thriving in the transition to AI-assisted software development}.
\newblock  (\bibinfo{year}{2024}).
\newblock


\bibitem[Holland and Wainer(2012)]%
        {holland2012differential}
\bibfield{author}{\bibinfo{person}{Paul~W Holland} {and} \bibinfo{person}{Howard Wainer}.} \bibinfo{year}{2012}\natexlab{}.
\newblock \bibinfo{booktitle}{\emph{Differential item functioning}}.
\newblock \bibinfo{publisher}{Routledge}.
\newblock


\bibitem[Jones(2000)]%
        {jones2000software}
\bibfield{author}{\bibinfo{person}{Capers Jones}.} \bibinfo{year}{2000}\natexlab{}.
\newblock \bibinfo{booktitle}{\emph{Software assessments, benchmarks, and best practices}}.
\newblock \bibinfo{publisher}{Addison-Wesley Longman Publishing Co., Inc.}
\newblock


\bibitem[Joshi et~al\mbox{.}(2023)]%
        {joshi2023tale}
\bibfield{author}{\bibinfo{person}{Spruha Joshi}, \bibinfo{person}{Samantha~M Doonan}, {and} \bibinfo{person}{John~R Pamplin~Ii}.} \bibinfo{year}{2023}\natexlab{}.
\newblock \showarticletitle{A tale of two cities: racialized arrests following decriminalization and recreational legalization of cannabis}.
\newblock \bibinfo{journal}{\emph{Drug and alcohol dependence}}  \bibinfo{volume}{249} (\bibinfo{year}{2023}), \bibinfo{pages}{109911}.
\newblock


\bibitem[Kaiser et~al\mbox{.}(2025)]%
        {kaiser2025new}
\bibfield{author}{\bibinfo{person}{Carolin Kaiser}, \bibinfo{person}{Jakob Kaiser}, \bibinfo{person}{Rene Schallner}, {and} \bibinfo{person}{Sabrina Schneider}.} \bibinfo{year}{2025}\natexlab{}.
\newblock \showarticletitle{A New Era of Online Search? A Large-Scale Study of User Behavior and Personal Preferences during Practical Search Tasks with Generative AI versus Traditional Search Engines}. In \bibinfo{booktitle}{\emph{Proceedings of the Extended Abstracts of the CHI Conference on Human Factors in Computing Systems}}. \bibinfo{pages}{1--7}.
\newblock


\bibitem[Kang(2021)]%
        {kang2021looking}
\bibfield{author}{\bibinfo{person}{Helen~H Kang}.} \bibinfo{year}{2021}\natexlab{}.
\newblock \showarticletitle{Looking toward restorative justice for redlined communities displaced by eco-gentrification}.
\newblock \bibinfo{journal}{\emph{Michigan Journal of Race and Law}}  \bibinfo{volume}{26} (\bibinfo{year}{2021}), \bibinfo{pages}{23--46}.
\newblock


\bibitem[Karnouskos(2024)]%
        {karnouskos2024relevance}
\bibfield{author}{\bibinfo{person}{Stamatis Karnouskos}.} \bibinfo{year}{2024}\natexlab{}.
\newblock \showarticletitle{The relevance of large Language models for project management}.
\newblock \bibinfo{journal}{\emph{IEEE Open Journal of the Industrial Electronics Society}}  \bibinfo{volume}{5} (\bibinfo{year}{2024}), \bibinfo{pages}{758--768}.
\newblock


\bibitem[Khan et~al\mbox{.}(2025)]%
        {khan2025randomness}
\bibfield{author}{\bibinfo{person}{Ariba Khan}, \bibinfo{person}{Stephen Casper}, {and} \bibinfo{person}{Dylan Hadfield-Menell}.} \bibinfo{year}{2025}\natexlab{}.
\newblock \showarticletitle{Randomness, not representation: The unreliability of evaluating cultural alignment in llms}. In \bibinfo{booktitle}{\emph{Proceedings of the 2025 ACM Conference on Fairness, Accountability, and Transparency}}. \bibinfo{pages}{2151--2165}.
\newblock


\bibitem[King(2017)]%
        {king2017status}
\bibfield{author}{\bibinfo{person}{LaGarrett~J King}.} \bibinfo{year}{2017}\natexlab{}.
\newblock \showarticletitle{The status of Black history in US schools and society}.
\newblock \bibinfo{journal}{\emph{Social Education}} \bibinfo{volume}{81}, \bibinfo{number}{1} (\bibinfo{year}{2017}), \bibinfo{pages}{14--18}.
\newblock


\bibitem[King(2020)]%
        {king2020black}
\bibfield{author}{\bibinfo{person}{LaGarrett~J King}.} \bibinfo{year}{2020}\natexlab{}.
\newblock \showarticletitle{Black history is not American history: Toward a framework of Black historical consciousness}.
\newblock \bibinfo{journal}{\emph{Social Education}} \bibinfo{volume}{84}, \bibinfo{number}{6} (\bibinfo{year}{2020}), \bibinfo{pages}{335--341}.
\newblock


\bibitem[Klassen(2022)]%
        {klassen2022black}
\bibfield{author}{\bibinfo{person}{Shamika Klassen}.} \bibinfo{year}{2022}\natexlab{}.
\newblock \showarticletitle{Black Twitter is gold: why this online community is worthy of study and how to do so respectfully}.
\newblock \bibinfo{journal}{\emph{Interactions}} \bibinfo{volume}{29}, \bibinfo{number}{1} (\bibinfo{year}{2022}), \bibinfo{pages}{96--98}.
\newblock


\bibitem[Lee et~al\mbox{.}(2024)]%
        {lee-etal-2024-kornat}
\bibfield{author}{\bibinfo{person}{Jiyoung Lee}, \bibinfo{person}{Minwoo Kim}, \bibinfo{person}{Seungho Kim}, \bibinfo{person}{Junghwan Kim}, \bibinfo{person}{Seunghyun Won}, \bibinfo{person}{Hwaran Lee}, {and} \bibinfo{person}{Edward Choi}.} \bibinfo{year}{2024}\natexlab{}.
\newblock \showarticletitle{{K}or{NAT}: {LLM} Alignment Benchmark for {K}orean Social Values and Common Knowledge}. In \bibinfo{booktitle}{\emph{Findings of the Association for Computational Linguistics: ACL 2024}}, \bibfield{editor}{\bibinfo{person}{Lun-Wei Ku}, \bibinfo{person}{Andre Martins}, {and} \bibinfo{person}{Vivek Srikumar}} (Eds.). \bibinfo{publisher}{Association for Computational Linguistics}, \bibinfo{address}{Bangkok, Thailand}, \bibinfo{pages}{11177--11213}.
\newblock
\href{https://doi.org/10.18653/v1/2024.findings-acl.666}{doi:\nolinkurl{10.18653/v1/2024.findings-acl.666}}


\bibitem[Li et~al\mbox{.}(2025)]%
        {li2025large}
\bibfield{author}{\bibinfo{person}{Meng Li}, \bibinfo{person}{Lai Wei}, {and} \bibinfo{person}{Yao Yao}.} \bibinfo{year}{2025}\natexlab{}.
\newblock \showarticletitle{Large Language Models in the Workplace: A Social Perspective}.
\newblock \bibinfo{journal}{\emph{Available at SSRN 5222006}} (\bibinfo{year}{2025}).
\newblock


\bibitem[McFarland(2024)]%
        {mcfarland2024cultural}
\bibfield{author}{\bibinfo{person}{Jon McFarland}.} \bibinfo{year}{2024}\natexlab{}.
\newblock \showarticletitle{Cultural Inclusivity and Gamification}.
\newblock \bibinfo{journal}{\emph{Pursuing Practical Change: Lesson Designs That Promote Culturally Responsive Teaching}} (\bibinfo{year}{2024}), \bibinfo{pages}{151}.
\newblock


\bibitem[Ponzini(2022)]%
        {ponzini2022public}
\bibfield{author}{\bibinfo{person}{Robert Ponzini}.} \bibinfo{year}{2022}\natexlab{}.
\newblock \showarticletitle{Public school funding in the United States and its systemic inequities}.
\newblock \bibinfo{journal}{\emph{Economia Aziendale Online-}} \bibinfo{volume}{13}, \bibinfo{number}{1} (\bibinfo{year}{2022}), \bibinfo{pages}{143--147}.
\newblock


\bibitem[Reuel et~al\mbox{.}(2024)]%
        {reuel2024betterbench}
\bibfield{author}{\bibinfo{person}{Anka Reuel}, \bibinfo{person}{Amelia Hardy}, \bibinfo{person}{Chandler Smith}, \bibinfo{person}{Max Lamparth}, \bibinfo{person}{Malcolm Hardy}, {and} \bibinfo{person}{Mykel~J Kochenderfer}.} \bibinfo{year}{2024}\natexlab{}.
\newblock \showarticletitle{Betterbench: Assessing ai benchmarks, uncovering issues, and establishing best practices}.
\newblock \bibinfo{journal}{\emph{Advances in Neural Information Processing Systems}}  \bibinfo{volume}{37} (\bibinfo{year}{2024}), \bibinfo{pages}{21763--21813}.
\newblock


\bibitem[Rodr{\'\i}guez and Swalwell(2023)]%
        {rodriguez2023social}
\bibfield{author}{\bibinfo{person}{Noreen~Naseem Rodr{\'\i}guez} {and} \bibinfo{person}{Katy Swalwell}.} \bibinfo{year}{2023}\natexlab{}.
\newblock \bibinfo{booktitle}{\emph{Social studies for a better world: An anti-oppressive approach for elementary educators}}.
\newblock \bibinfo{publisher}{Routledge}.
\newblock


\bibitem[Rohovyi and Grinchenko(2024)]%
        {rohovyi2024towards}
\bibfield{author}{\bibinfo{person}{Mykyta Rohovyi} {and} \bibinfo{person}{Marina Grinchenko}.} \bibinfo{year}{2024}\natexlab{}.
\newblock \showarticletitle{Towards the improvement of project team performance based on large language models}.
\newblock \bibinfo{journal}{\emph{Radioelectronic and Computer Systems}} \bibinfo{volume}{2024}, \bibinfo{number}{4} (\bibinfo{year}{2024}), \bibinfo{pages}{229--247}.
\newblock


\bibitem[Said et~al\mbox{.}(2013)]%
        {said2013benchmarking}
\bibfield{author}{\bibinfo{person}{Alan Said}, \bibinfo{person}{Domonkos Tikk}, {and} \bibinfo{person}{Paolo Cremonesi}.} \bibinfo{year}{2013}\natexlab{}.
\newblock \showarticletitle{Benchmarking: A methodology for ensuring the relative quality of recommendation systems in software engineering}.
\newblock In \bibinfo{booktitle}{\emph{Recommendation Systems in Software Engineering}}. \bibinfo{publisher}{Springer}, \bibinfo{pages}{275--300}.
\newblock


\bibitem[Tao et~al\mbox{.}(2024)]%
        {tao2024cultural}
\bibfield{author}{\bibinfo{person}{Yan Tao}, \bibinfo{person}{Olga Viberg}, \bibinfo{person}{Ryan~S Baker}, {and} \bibinfo{person}{Ren{\'e}~F Kizilcec}.} \bibinfo{year}{2024}\natexlab{}.
\newblock \showarticletitle{Cultural bias and cultural alignment of large language models}.
\newblock \bibinfo{journal}{\emph{PNAS nexus}} \bibinfo{volume}{3}, \bibinfo{number}{9} (\bibinfo{year}{2024}), \bibinfo{pages}{pgae346}.
\newblock


\bibitem[Wang et~al\mbox{.}(2025)]%
        {wang2025generative}
\bibfield{author}{\bibinfo{person}{Tianjia Wang}, \bibinfo{person}{Tong Wu}, \bibinfo{person}{Huayi Liu}, \bibinfo{person}{Chris Brown}, {and} \bibinfo{person}{Yan Chen}.} \bibinfo{year}{2025}\natexlab{}.
\newblock \showarticletitle{Generative Co-Learners: Enhancing Cognitive and Social Presence of Students in Asynchronous Learning with Generative AI}.
\newblock \bibinfo{journal}{\emph{Proceedings of the ACM on Human-Computer Interaction}} \bibinfo{volume}{9}, \bibinfo{number}{1} (\bibinfo{year}{2025}), \bibinfo{pages}{1--24}.
\newblock


\bibitem[Williams(2020)]%
        {williams2020fitting}
\bibfield{author}{\bibinfo{person}{Damien~Patrick Williams}.} \bibinfo{year}{2020}\natexlab{}.
\newblock \showarticletitle{Fitting the description: Historical and sociotechnical elements of facial recognition and anti-black surveillance}.
\newblock \bibinfo{journal}{\emph{Journal of Responsible Innovation}} \bibinfo{volume}{7}, \bibinfo{number}{sup1} (\bibinfo{year}{2020}), \bibinfo{pages}{74--83}.
\newblock


\end{thebibliography}

\end{document}